\documentclass[dvipsnames, twocolumn]{aastex631}

\usepackage{amsmath}
\usepackage{graphicx}
\usepackage{enumitem}

\newcommand{\ee}[1]{\mbox{${} \times 10^{#1}$}}
\newcommand{\lsun}{\mbox{L$_\odot$}}
\newcommand{\msun}{\mbox{M$_\odot$}}
\newcommand{\as}{\mbox{\arcsec}}
\newcommand{\degree}{\mbox{$^{\circ}$}}

\newcommand{\um}{$\mathrm{\mu}$m}

\newcommand{\kms}{km~\,s$^{-1}$}

\newcommand{\hh}{\mbox{{\rm H}$_2$}}

\def\jwst{\emph{JWST}}
\def\spitzer{\emph{Spitzer}}

\begin{document}

\title{MINDS. JWST-MIRI Observations of a Spatially Resolved Atomic Jet and Polychromatic Molecular Wind Toward SY Cha}

\author[0000-0002-6429-9457]{Kamber R. Schwarz}
\affil{Max-Planck-Institut f\"{u}r Astronomie (MPIA), K\"{o}nigstuhl 17, 69117 Heidelberg, Germany}

\author[0000-0001-9992-4067]{Matthias Samland}
\affil{Max-Planck-Institut f\"{u}r Astronomie (MPIA), K\"{o}nigstuhl 17, 69117 Heidelberg, Germany}

\author[0000-0003-3747-7120]{G\"oran Olofsson}
\affil{Department of Astronomy, Stockholm University, AlbaNova University Center, 10691 Stockholm, Sweden}

\author[0000-0002-1493-300X]{Thomas Henning}
\affil{Max-Planck-Institut f\"{u}r Astronomie (MPIA), K\"{o}nigstuhl 17, 69117 Heidelberg, Germany}

\author[0000-0003-0330-1506]{Andrew Sellek}
\affil{Leiden Observatory, Leiden University, PO Box 9513, 2300 RA Leiden, the Netherlands}

\author[0000-0001-9818-0588]{Manuel G\"udel}
\affil{Dept. of Astrophysics, University of Vienna, T\"urkenschanzstr. 17, A-1180 Vienna, Austria}
\affil{ETH Z\"urich, Institute for Particle Physics and Astrophysics, Wolfgang-Pauli-Str. 27, 8093 Z\"urich, Switzerland}

\author[0000-0002-1103-3225]{Beno\^{i}t Tabone}
\affil{Universit\'e Paris-Saclay, CNRS, Institut d’Astrophysique Spatiale, 91405, Orsay, France}


\author[0000-0001-7455-5349]{Inga Kamp}
\affil{Kapteyn Astronomical Institute, Rijksuniversiteit Groningen, Postbus 800, 9700AV Groningen, The Netherlands}

\author{Pierre-Olivier Lagage}
\affil{Universit\'e Paris-Saclay, Universit\'e Paris Cit\'e, CEA, CNRS, AIM, F-91191 Gif-sur-Yvette, France}

\author[0000-0001-7591-1907]{Ewine F. van Dishoeck}
\affil{Leiden Observatory, Leiden University, PO Box 9513, 2300 RA Leiden, the Netherlands}
\affil{Max-Planck Institut f\"{u}r Extraterrestrische Physik (MPE), Giessenbachstr. 1, 85748, Garching, Germany}


\author[0000-0001-8876-6614]{Alessio Caratti o Garatti}
\affil{INAF – Osservatorio Astronomico di Capodimonte, Salita Moiariello 16, 80131 Napoli, Italy}
\affil{Dublin Institute for Advanced Studies, 31 Fitzwilliam Place, D02 XF86 Dublin, Ireland}

\author[0000-0001-9250-1547]{Adrian M. Glauser}
\affil{ETH Z\"urich, Institute for Particle Physics and Astrophysics, Wolfgang-Pauli-Str. 27, 8093 Z\"urich, Switzerland}

\author[0000-0002-2110-1068]{Tom P. Ray}
\affil{Dublin Institute for Advanced Studies, 31 Fitzwilliam Place, D02 XF86 Dublin, Ireland}


\author[0000-0001-8407-4020]{Aditya M. Arabhavi}
\affil{Kapteyn Astronomical Institute, Rijksuniversiteit Groningen, Postbus 800, 9700AV Groningen, The Netherlands}

\author[0000-0002-0101-8814]{Valentin Christiaens}
\affil{STAR Institute, Universit\'e de Li\`ege, All\'ee du Six Ao\^ut 19c, 4000 Li\`ege, Belgium}

\author[0000-0002-8889-2992]{R. Franceschi}
\affil{Max-Planck-Institut f\"{u}r Astronomie (MPIA), K\"{o}nigstuhl 17, 69117 Heidelberg, Germany}

\author[0000-0002-1257-7742]{Danny Gasman}
\affil{Institute of Astronomy, KU Leuven, Celestijnenlaan 200D, 3001 Leuven, Belgium}

\author[0000-0002-4022-4899]{Sierra L. Grant}
\affil{Max-Planck Institut f\"{u}r Extraterrestrische Physik (MPE), Giessenbachstr. 1, 85748, Garching, Germany}

\author[0000-0003-0386-2178]{Jayatee Kanwar}
\affil{Kapteyn Astronomical Institute, Rijksuniversiteit Groningen, Postbus 800, 9700AV Groningen, The Netherlands}
\affil{Space Research Institute, Austrian Academy of Sciences, Schmiedlstr. 6, A-8042, Graz, Austria}
\affil{TU Graz, Fakultät für Mathematik, Physik und Geodäsie, Petersgasse 16 8010 Graz, Austria}

\author[0000-0001-8240-978X]{Till Kaeufer}
\affil{Space Research Institute, Austrian Academy of Sciences, Schmiedlstr. 6, A-8042, Graz, Austria}
\affil{Kapteyn Astronomical Institute, Rijksuniversiteit Groningen, Postbus 800, 9700AV Groningen, The Netherlands}
\affil{SRON Netherlands Institute for Space Research, Niels Bohrweg 4, NL-2333 CA Leiden, the Netherlands}
\affil{Institute for Theoretical Physics and Computational Physics, Graz University of Technology, Petersgasse 16, 8010 Graz, Austria}

\author[0000-0002-2358-4796]{Nicolas T. Kurtovic}
\affil{Max-Planck Institut f\"{u}r Extraterrestrische Physik (MPE), Giessenbachstr. 1, 85748, Garching, Germany}

\author[0000-0002-8545-6175]{Giulia Perotti}
\affil{Max-Planck-Institut f\"{u}r Astronomie (MPIA), K\"{o}nigstuhl 17, 69117 Heidelberg, Germany}

\author[0000-0002-7935-7445]{Milou Temmink}
\affil{Leiden Observatory, Leiden University, PO Box 9513, 2300 RA Leiden, the Netherlands}

\author[0000-0002-3135-2477]{Marissa Vlasblom}
\affil{Leiden Observatory, Leiden University, PO Box 9513, 2300 RA Leiden, the Netherlands}

\begin{abstract}
The removal of angular momentum from protostellar systems drives accretion onto the central star and may drive the dispersal of the protoplanetary disk. Winds and jets can contribute to removing angular momentum from the disk, though the dominant process remain unclear. To date, observational studies of resolved disk winds have mostly targeted highly inclined disks. We report the detection of extended \hh\ and [Ne II] emission toward the young stellar object SY Cha with the \emph{JWST} Mid-InfraRed Instrument Medium Resolution Spectrometer (MIRI-MRS). This is one of the first polychromatic detections of extended \hh\ toward a moderately inclined, $i=51.1\degree$, Class II source. 
We measure the semi-opening angle of the \hh\ emission as well as build a rotation diagram to determine the \hh\ excitation temperature and abundance.  
We find a wide semi-opening angle, high temperature, and low column density for the \hh\ emission, all of which are characteristic of a disk wind. We derive a molecular wind mass loss rate of $3\pm2\ee{-9}$ \msun\ yr$^\mathrm{-1}$, which is high compared to the previously derived stellar accretion rate of $6.6\ee{-10}$ \msun\ yr$^\mathrm{-1}$. This suggests either that the stellar accretion and the disk wind are driven by different mechanisms or that accretion onto the star is highly variable.
These observations demonstrate MIRI-MRS's utility in expanding studies of resolved disk winds beyond edge-on sources.

\end{abstract}

\section{Introduction} \label{sec:intro}
Stars form from collapsing clouds of gas and dust. Due to conservation of angular moment, not all of the material is able to collapse directly onto the star, instead forming a protoplanetary disk. In order for material to accrete from the disk onto the star it must first shed its angular momentum. 
Turbulence driven viscous spreading can redistribute angular momentum within the disk \citep{Shakura73}. Alternatively, outflows can remove angular momentum from the system entirely \citep{Pudritz83}. Outflows also drive the dissipation of the protoplanetary disk. 

Outflows are variously classified as winds and jets based on the degree of collimation and outflow velocity. Winds launched by the disk can be driven either by high-energy radiation from the central star: photoevaporative winds, or by the stellar and/or disk magnetic field: magnetohydrodynamic (MHD) winds; see \citet{Pascucci23} for a recent review. Photoevaporative winds are expected to be launched from radii greater than a tenth of the gravitational radius but small enough for the disk to intercept a large amount of high energy radiation from the central star, and to be relatively slow, with velocities $\leq$ 10 km~s$^{-1}$ \citep{Hollenbach94,Liffman03,Clarke16}. In contrast, MHD winds can be launched from a large range of radii, and have larger initial velocities, and thus larger vertical extents \citep{Blandford82, Lesur21}. Which of these processes is the dominant form of mass removal remains unclear \citep{Hartmann16}. Jets are both more collimated (semi-opening angle $< 20\degree$) and have a higher velocity ($\geq 100$ \kms) than winds \citep{Frank14}. They are launched via MHD processes in the inner most regions of the disk or at the stellar surface, though the exact details are still uncertain \citep{Tsukamoto23}.

Forbidden line emission from atomic ions can trace both winds and jets \citep{Pascucci07,vanBoekel09,Pascucci20,Flores23,Tychoniec24,Bajaj24,Federman24}. 
Extended emission from ro-vibrational as well as pure rotation lines of \hh\ has been observed toward a number of young stellar objects (YSOs) and is often attributed to either an MHD disk wind or shocks in the outflow cavity \citep{Beck08,Melnikov23,Arulanantham24,Delabrosse24,Tychoniec24}.

Here we present multi-line analysis of extended \hh\ emission toward a Class II protoplanetary disk. SY Cha hosts a moderately inclined transition disk, displaying a 70 au cavity seen both in millimeter continuum emission \citep{Orihara23} and NIR scattered light \citep{Ren23,Juillard24,Ginski24}. A summary of the stellar and disk properties are given in Table~\ref{sourcetab}. 
The \emph{JWST} observations were taken as part of the MIRI mid-Infrared Disk Survey (MINDS). For a description of the full sample and observing strategy see \citet{Kamp23} and \citet{Henning24}.

\begin{deluxetable}{lll}
\tablewidth{\columnwidth}
\tablecolumns{3}
\tablecaption{Properties of SY Cha}
\label{sourcetab}
\tablehead{
\colhead{Parameter} & \colhead{Value}  & \colhead{Refs}  
}
\startdata
Stellar mass & 0.70 \msun\ & 1,2,3 \\
Stellar luminosity & 0.55 \lsun\ & 1,2,3 \\
Effective temperature & 4030 K & 1,2,3 \\
Accretion luminosity & 7.39\ee{-3} \lsun\ yr$^{-1}$ & 1,2,3 \\
Mass accretion rate & 6.6\ee{-10} \msun\ yr$^{-1}$ & 1,2,3 \\
Spectral type & K5Ve & 4\\
A$_\mathrm{v}$ & 0.5 & 5 \\
Distance & 180.7 pc & 3 \\
Disk inclination & 51.1\degree\ & 6 \\
\enddata
\tablecomments{
1 = \citet{Manara16}, 2 = \citet{Feiden16}, 3 = \citet{Gaia21}, 4 = \citet{Frasca15}, 5 = \citep{Manara16}, 6 = \citet{Orihara23}
}
\end{deluxetable}

\section{Observations and Data Reduction} \label{reduction}
Observations of SY Cha with the Mid-InfraRed Instrument (MIRI) \citet{Rieke15,Wright15} in Medium Resolution Spectroscopy (MRS) \citet{Wells15} mode were carried out on 8 August, 2022 as part of the MINDS GTO Program (PID: 1282, PI: T. Henning). As the source is bright, dedicated target acquisition was not carried out. A four-point dither patter in the positive direction was used and the total exposure time was 3696 seconds. 
Data reduction was carried out following the process described by \citet{Schwarz24}, though with an updated version (1.13.3) of the \emph{JWST} pipeline \citep{Bushouse23}. 
Visual inspection of the resulting data cubes reveals extended emission at wavelengths corresponding to the \hh\ and [Ne~II] lines. Here we focus on this extended emission. For information on the spatially unresolved continuum and line emission, see \citet{Schwarz24}.

\begin{figure*}
    \centering
    \includegraphics[width=\textwidth]{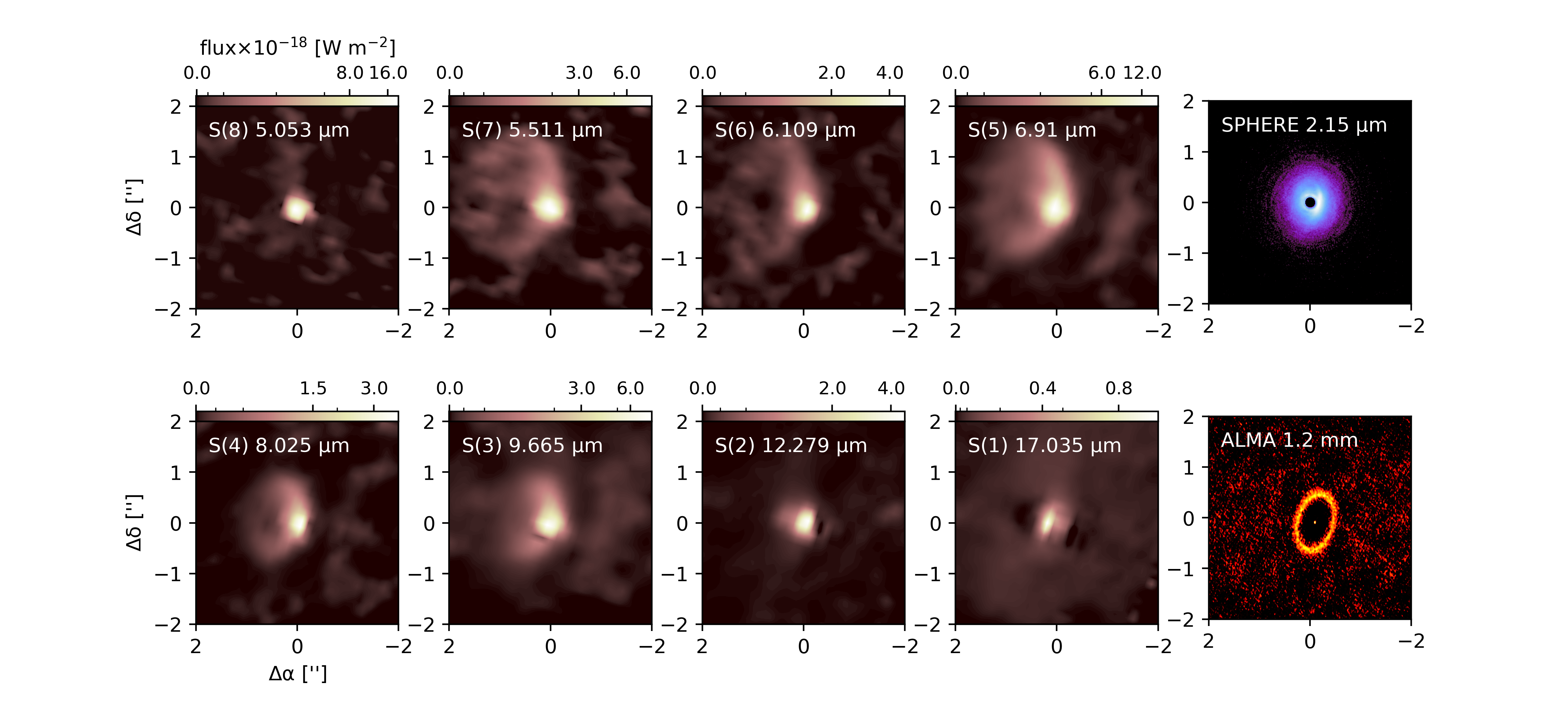}
    \caption{\hh\ lines after nine rounds of de-convolution. The right panels show the scattered light emission observed by SPHERE and imaged by \citet{Juillard24} and the dust continuum emission observed with ALMA \citep{Orihara23}.}
    \label{fig:itr9}
\end{figure*}

\begin{figure}
    \centering
    \includegraphics[width=\columnwidth]{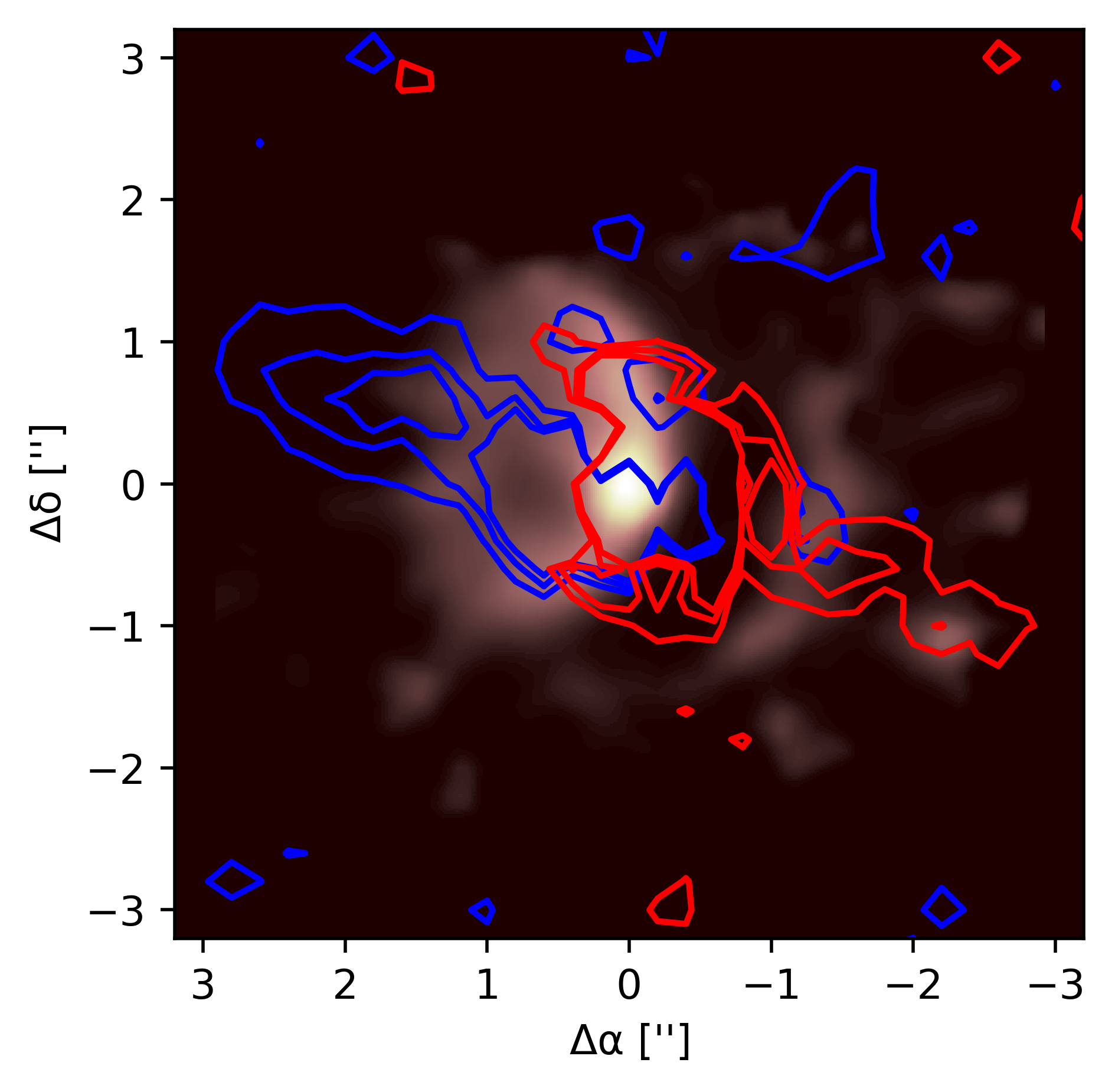}
    \caption{Blue and red shifted wings of the [Ne II] emission, with an offset of 117 km/s. Background is the $S$(5) \hh\ line.}
    \label{fig:NeII}
\end{figure}

The extended \hh\ emission is not easily seen in the spectral images directly because it is comparable in surface brightness to the wings of the PSF (Point Spread Function). However, after the subtraction of the adjacent continuum spectral images the extended emission is revealed (see Appendix~\ref{app:deconvolution}), indicating an outflow structure. In order to sharpen these \hh\ images we apply a de-convolution method, the Lucy-Richardson algorithm \citep{Richardson72,Lucy74}. This requires the PSF for each \hh\ line but instead of using theoretical templates or observations of calibration point sources we simply use the continuum spectral images. The advantage is the identical dithering and preprocessing, although we cannot be sure that the source is point-like in the continuum. The number of adjacent spectral images is a trade-off between the signal-to-noise ratio (S/N) and how closely the deduced `PSF' will represent that of the \hh\ line wavelength. As a compromise we chose the median of 30 spectral images at each side of the emission line which is represented by the co-added 5 spectral images covering the line.
 By increasing the number of iterations, the features get sharper (Appendix~\ref{app:deconvolution}) and how far one can proceed depends basically on the S/N ratio. The observed \hh\ lines differ in this respect and as a compromise we chose 9 iterations (Figure~\ref{fig:itr9}). These images are used in all following analysis unless otherwise noted. The $S$(3)-$S$(7) lines all appear spatially extended. 
The [Ne II] emission also appears spatially extended (Figure~\ref{fig:NeII}).
 The excess blue- and red-shifted [Ne II] emission after PSF and continuum subtraction is shown in Figure~\ref{fig:NeII}. Deconvolution is not performed on the [Ne II] emission.  [Ne III] and [Ar II] are detected but do not appear extended. No other forbidden lines are detected. Notably, there is no evidence of the [Fe II] line at 17.936~\micron, though a weak feature could be masked by overlapping water emission \citep{Schwarz24}.

\section{Analysis}\label{analysis}

The [Ne II] emission originates primarily from the region around the central source. Additionally, extended emission is clearly seen on either side of the central source, roughly perpendicular to the protoplanetary disk \citep{Orihara23,Ginski24}. This extended emission is strongest in the spectral images adjacent to the central [Ne II] frame at 12.814 \um (see Appendix~\ref{app:channel}). The blue- and red-shifted emission is binned with velocity offset of 117~km~s$^{-1}$, extending roughly 3\as\ from the central source in opposite directions. The blue-shifted emission appears brighter, as expected for a bipolar outflow where the light from the red-shifted lobe is subject to more extinction along the line of sight. Following the method of \citet{Pascucci24}, we determine the semi-opening angle of the blue and red-shifted [NeII] emission by fitting a Gaussian profile to cuts perpendicular to the outflow axis. This allows us to derive a FWHM. At this wavelength the pixel scale is $0.2\as$. Any derived FWHM below this value is considered unresolved and not used to determine the semi-opening angle. We then derive the semi-opening angle from the slope of line fit to the remaining data points. The derived semi-opening angle for the blue-shifted lobe is $6.3\pm1.8\degree$, consistent with measurements for jets in other sources \citep{Delabrosse24,Arulanantham24,Pascucci24}.

The \hh\ $S(3)-S(7)$ extended emission primarily originates from the near side of the disk, the same side as the blue-shifted [Ne II]. There are also hints of extended emission on the far side of disk, most clearly seen in the $S$(5) line. The \hh\ emission appears less collimated than the [Ne II] emission. The higher energy transitions generally appear more extended than the lower energy transitions: the $3\sigma$ contour for the $S$(7) in Figure~\ref{fig:itr9} extends $1.78\as$ perpendicular to the plane of the protoplanetary disk and $1.35\as$ horizontally, while for the $S$(3) it is only $0.99\as$ and $0.96\as$.
This difference could be due to the lower signal to noise at longer wavelengths. 

\begin{figure*}
    \centering
    \includegraphics[width=0.9\textwidth]{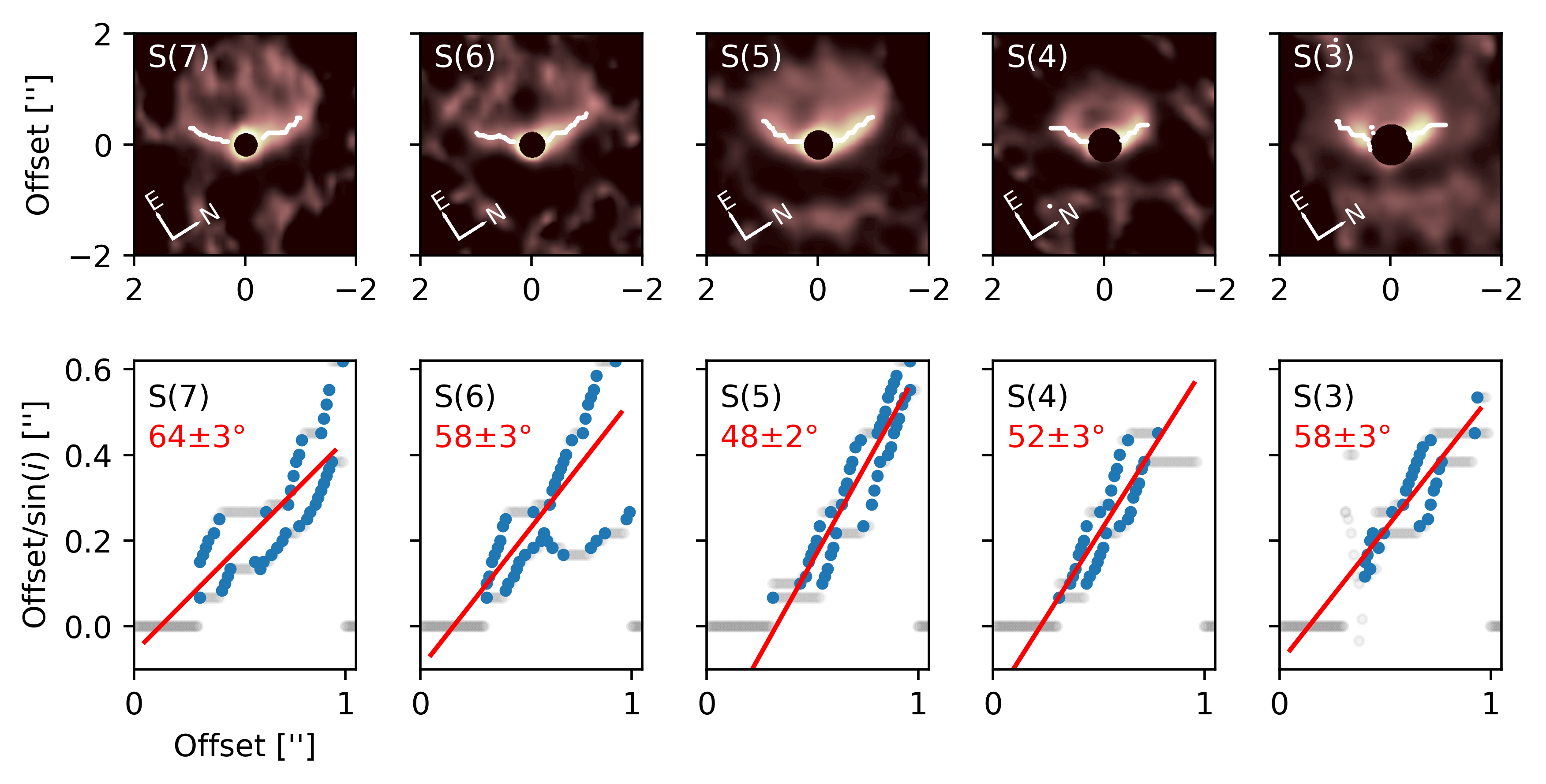}
    \caption{Top: White points indicate the brightest pixel at each position along the x-axis for the five \hh\ lines with extended emission. The central region inside the diffraction limit is masked to better show the faint extended emission. Bottom: Best fit line (red) tracing the extended emission. For a given vertical offset only the point with the smallest horizontal offset is used in the fit (blue points). Points not used in the fit are shown in grey. The derived semi-opening angle for each line is listed in red.}
    \label{fig:openang}
\end{figure*}

To determine the semi-opening angle of the \hh\ emission we first identify the offset of the brightest pixel at each position along the x-axis (Figure~\ref{fig:openang}). This vertical offset is de-projected assuming the inclination of the \hh\ emission is the same as the disk inclination. We then overlay the two sides before fitting a straight line and deriving the semi-opening angle from our fit, as seen in the bottom panel of Figure~\ref{fig:openang}. 
The derived semi-opening angles range from $64\pm 2\degree$ for $S$(7) to $48\pm 1\degree$ for $S$(5). It should be noted that pixel scale of the MRS increases from 0.196\as\ in Channel 1 to 0.273\as\ in Channel 4. Additionally, the full width half max (FWHM) of the PSF 
is known to increase with wavelength \citep{Argyriou23,Law23}.  

\subsection{Molecular Hydrogen Rotation Diagram}
The \hh\ emission could potentially originate from a wind or from the hot surface of the disk. 
For a population of molecules in LTE, the number of molecules at each upper state excitation level, $N_u$, is given by
\begin{equation}\label{eq:Nu}
N_u = \frac{N}{Z} g_u \exp{\left(-E_u /k T\right)}
\end{equation}
where $N$ is the total column density, $Z(T)$ is the partition function for a given excitation temperature $T$, $g_u$ and $E_u$ are the statistical weight and upper state energy, and $k$ is the Boltzmann constant. 
Assuming optically thin emission, the observed flux $F_u$ is related to the upper state column density:
\begin{equation}
N_u = \frac{4\pi F_u}{A_u h \nu \Omega}
\end{equation}
where $\Omega$ is the emitting area and $A_u$ is the Einstein A rate coefficient for spontaneous emission. Equation~\ref{eq:Nu} can then be rewritten as:
\begin{equation}
\ln{ \frac{N_u}{ g_u}} = \ln{ \frac{4\pi F_u}{g_u A_u h \nu \Omega}} = -\frac{E_u}{k T} - \ln{\frac{Z(T)}{N}}
\end{equation}

We extract the flux for each line by summing the flux within 0.5\as\ and 2\as\ radius apertures centered on the source, allowing us to differentiate between emission at the central source and extended emission. The values used in constructing the rotation diagram are given in Table~\ref{tab:h2flux}, where the uncertainty is propagated from the rms noise and added in quadrature with the $5\%$ absolute flux calibration uncertainty \citep{Argyriou23}. We do not correct correct for extinction along the line of sight as the visual extinction toward SY Cha is low \citep[A$\mathrm{_V}$ of 0.5][]{Manara16} and the uncertainty in converting to an IR optical depth is large \citep{McClure09}.

We determine the column density $N$ by using our best fit excitation temperature $T$, to solve for the partition function $Z$. Here we use the normal \hh\ partition function of \citet{Popovas16}, which treats the ortho and para \hh\ separately. We convert the derived column density into a total \hh\ mass assuming that the emitting area is equal to our aperture. $2\as$ is an overestimate of the true emitting area and thus results in an upper limit on the true \hh\ mass. The resulting values are given in Table~\ref{tab:rotdiag}.


\begin{deluxetable*}{lccccccc}
\tablecolumns{4}
\tablecaption{\hh\ Line Data}
\label{tab:h2flux}
\tablehead{
\multicolumn{4}{c}{}  & \multicolumn{3}{c}{Flux Density} \\
\colhead{Transition} & \colhead{Line Center} & \colhead{$E_u$} & \colhead{$g_u$} & \colhead{$n_{crit}$} & \multicolumn{3}{c}{\ee{-4} [erg s$^{-1}$ cm$^{-2}$ str$^{-1}$]}  \\
\colhead{} & \colhead{[\micron]} & \colhead{[K]} & \colhead{} & \colhead{[cm$^\mathrm{-3}$]} & \colhead{(0-2\as)} & \colhead{(0-0.5\as)} & \colhead{(0.5-2.0\as)}
}
\startdata
$S$(1)  &  17.0348  &  1015.1 & 21 & 2.4e2 & $0.74\pm0.04$  & $3.2\pm0.2$ & $0.58\pm0.05$ \\
$S$(2)  &  12.2786  &  1681.6 & 9  & 1.7e3 & $1.10 \pm0.06$ & $12.7\pm0.7$ & $0.33\pm0.08$ \\
$S$(3)  &  9.66491  &  2503.7 & 33 & 5.5e3 & $3.6\pm0.2$  & $27\pm 1$ & $2.0\pm0.2$ \\
$S$(4)  &  8.02504  &  3474.5 & 13 & 2.0e4 & $1.38 \pm0.08$ & $11.3\pm0.6$ & $0.7\pm0.1$ \\
$S$(5)  &  6.90951  &  4586.1 & 45 & 3.9e4 & $7.4\pm0.4$   & $60\pm3$ & $3.9\pm0.5$ \\
$S$(6)  &  6.10856  &  5829.8 & 17 & 1.0e5 & $2.2\pm 0.1$ & $16.6\pm0.9$ & $1.2\pm0.1$ \\
$S$(7)  &  5.51118  &  7196.7 & 57 & 1.5e5 & $4.4\pm0.3$  & $33\pm2$ & $2.5\pm0.3$ \\
$S$(8)  &  5.05306  &  8677.1 & 21 & \nodata\ & $2.6\pm0.1$  & $30\pm2$ & $0.7\pm0.2$ \\
\enddata
\tablecomments{
$E_{u}$ and $g_{u}$  values from \citet{Roueff19}. $n_{crit}$ for collisions with H and a temperature of 1000~K from \citet{LeBourlot99}
}
\end{deluxetable*}

\begin{figure*}[!h]
    \centering
    \plottwo{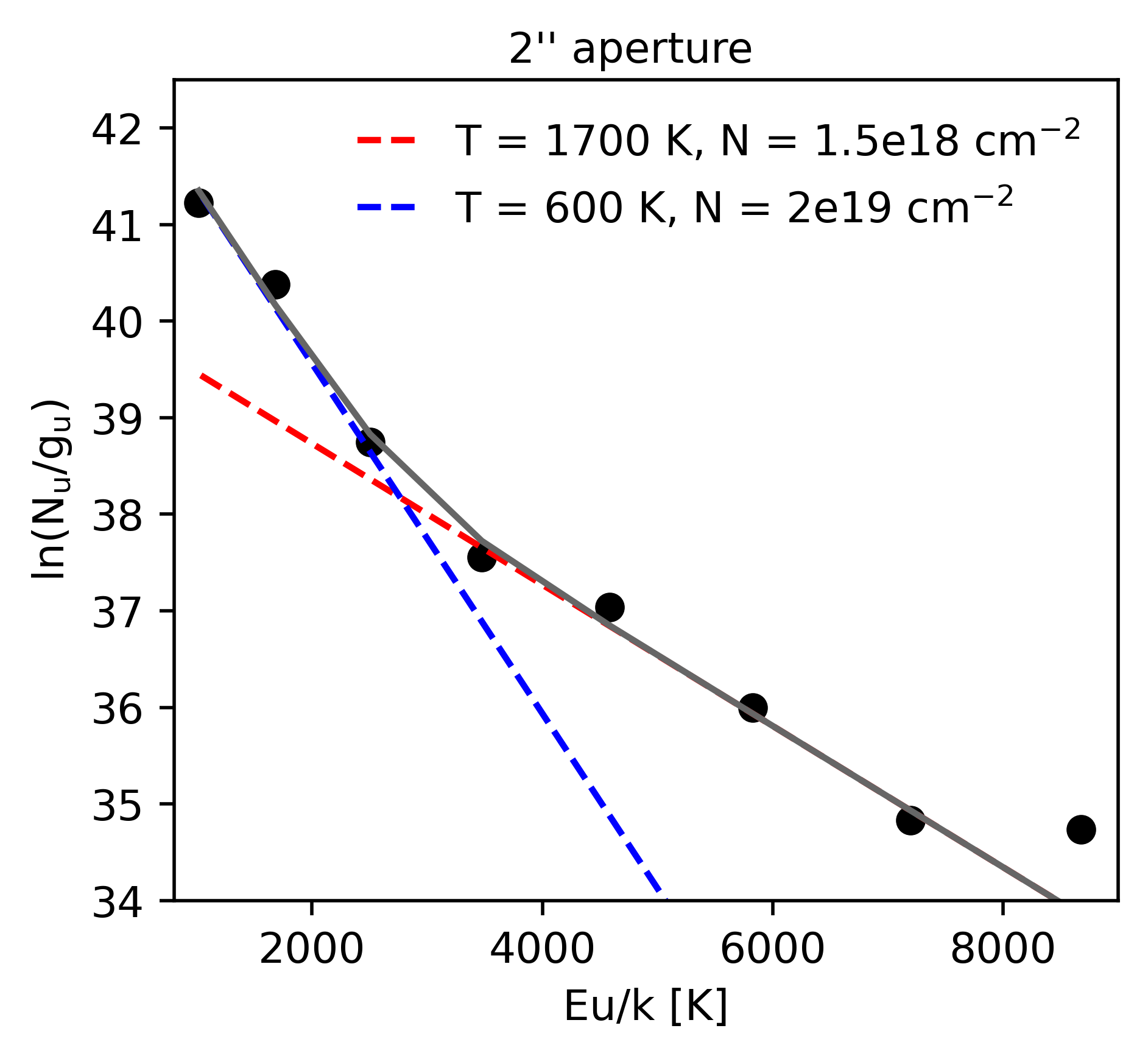}{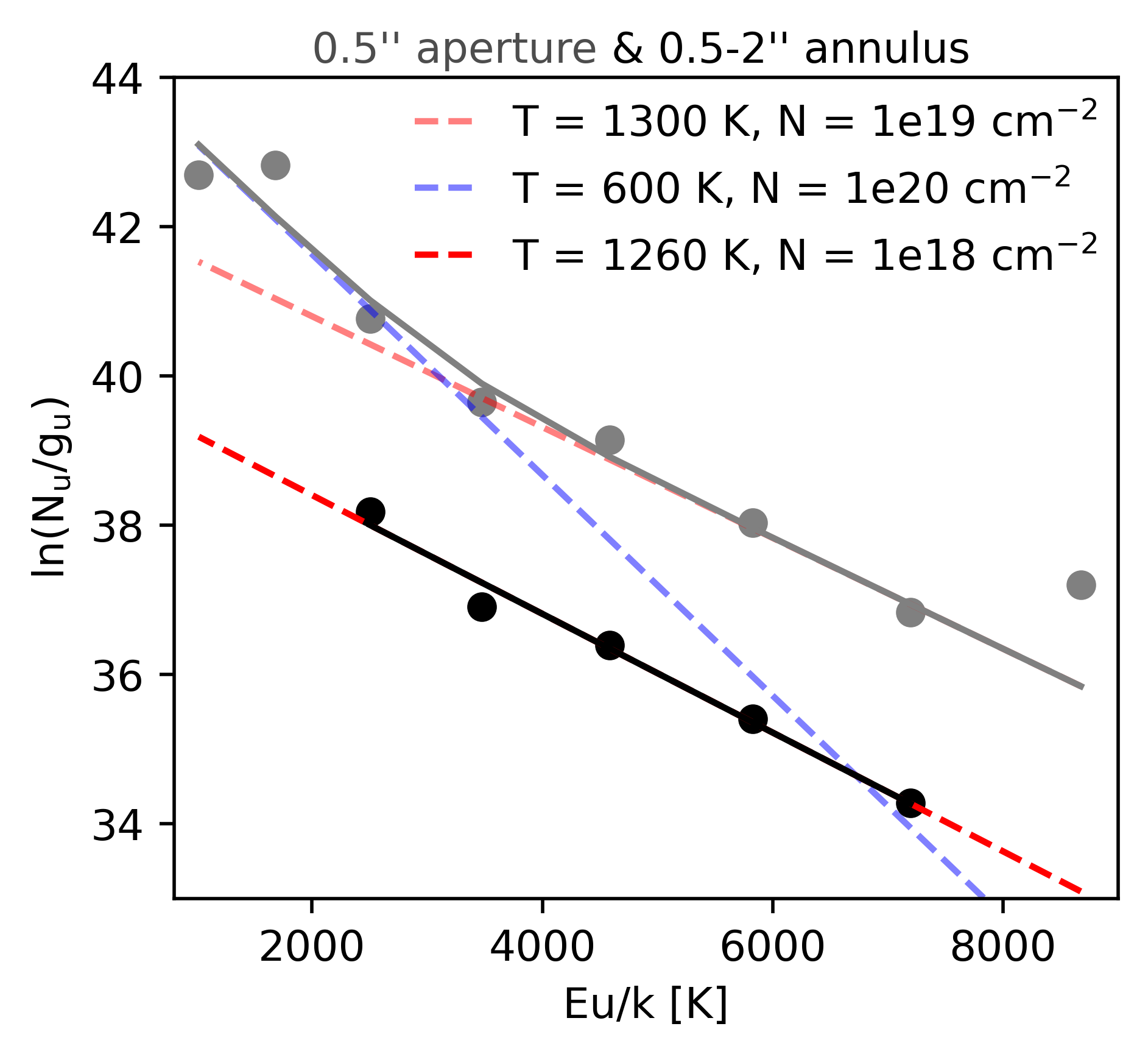}
    \caption{Rotation diagrams with a two component fit. The left panel shows the fit for all flux within a 2\as\ radius aperature centered on the source. The right panel show the fit for the flux within a 0.5\as\ aperture (grey) and the flux within an annulus between 0.5 and 2.0\as (black).}
    \label{rotdiag}
\end{figure*}

There is a clear change of slope in the rotation diagram in Figure~\ref{rotdiag}, necessitating a two component fit. The hot component is fit by a temperature of $1700\pm 300$~K and a column density of $6.0  \pm  0.1 \ee{19}$~cm$^{-2}$.
The warm component is fit by a temperature  of $600\pm 100$~K, consistent with emission originating from the disk surface \citep[e.g.,][]{Franceschi24}, and a column density of $2  \pm  1 \ee{19}$~cm$^{-2}$ (Table~\ref{tab:rotdiag}). 
It is also clear from Figure~\ref{rotdiag} that the lines for which we detect extended emission ($E_u=2503.7$ to 7196.7 K) have a large contribution from the hot component. To test if the two temperature components are due to two distinct origins, rather than a gradual temperature gradient, we redo our analysis on only the emission within the inner 0.5\as\ and on only the emission from a annulus from 0.5 to 2\as\ (Figure~\ref{rotdiag}, right panel). For the inner region, we exclude the 8677.1~K point from the fit, as this results in an unrealistically high temperature for the hot component. As for the full 2\as\ region, fitting the inner 0.5\as\ requires a two component fit. For the outer annulus, we only consider those lines for which extended emission is observed. For this region a single component fit is sufficient. $T_{warm}$ for the inner 0.5\as\ is the same as for the full region. $T_{hot}$ hot in inner and outer regions agree within the uncertainty, $~1300$~K , though both values are several hundred Kelvin lower compared to the full 2.0\as. The higher temperature for the full aperture measurement is due to the inclusion of the S(8) line. Excluding this line results in $\mathrm{T_{warm} = 1300\pm200}$ K. The overall consistency between the derived temperatures suggests two distinct origins: one, limited to the inner regions and characterized by a temperature of 600~K, likely tracing the hot disk surface, and one characterized by a temperature of 1300-1700~K, likely an outflow. 

We can use the derived column density for the hot component to check if our assumption of LTE is valid. To convert from an \hh\ column density to an average number density we divide the column density by the observed extent of the emission on the sky of 1.78\as. This results in an average \hh\ number density of $n(\hh) = 228$ cm$^\mathrm{-3}$. In this high temperature, low density regime is expected to be in form of neutral H or H$^+$, with n(H) more than a factor of 100 greater than n(\hh) \citep{Bron16,Nakatani18,Sellek24b}. This makes atomic hydrogen the main collisional partner with a density of $n(H) \sim 2\ee{4}$ cm$^\mathrm{-3}$.
According to the critical densities for collisions with H atoms at 1000~K given in Table~\ref{tab:h2flux}, only the lowest energy lines for which we observe extended emission are expected to be in LTE \citep{LeBourlot99}. LTE assumes the population distribution for energy levels of a given species can be described by a single excitation temperature. Using LTE assumptions for a population in non-LTE can lead to a more than order of magnitude increase in the uncertainty of the derived column density \citep{Mangum15}. As such, the column density reported here should be considered tentative. More advanced modeling based analysis is needed to obtain more precise measurements \citep[e.g,][]{vanderTak07,Brinch10}. 

\begin{deluxetable*}{lccc}
\tablewidth{\textwidth}
\tablecolumns{4}
\tablecaption{\hh\ Properties Derived from Rotation Diagram}
\label{tab:rotdiag}
\tablehead{
\colhead{Parameter} & \colhead{2.0\as} & \colhead{0.5\as} & \colhead{0.5-2.0\as}
}
\startdata
T$_\mathrm{hot}$ [K]& $1700 \pm 300$  & $1300 \pm 500$ &    $1260\pm90$ \\
T$_\mathrm{warm}$ [K]& $600 \pm 100$ & $600 \pm 300$   &    \nodata \\
N$_\mathrm{hot}$ [cm$^{-2}$] & $1.5 \pm 1.0 \ee{18}$ & $1 \pm 2 \ee{19}$  & $1.1\pm0.3 \ee{18}$ \\
N$_\mathrm{warm}$ [cm$^{-2}$] & $2.0 \pm 1.0 \ee{19}$ & $1 \pm 1\ee{20}$   &  \nodata \\
M$_\mathrm{hot}$(\hh) [\msun] & $4 \pm 3\ee{-7}$  & $2 \pm 4\ee{-7}$ & $3.1\pm0.9\ee{-7}$ \\
M$_\mathrm{warm}$(\hh) [\msun] & $7 \pm 4\ee{-6}$ & $2 \pm 2 \ee{-6}$ & \nodata \\
\enddata
\end{deluxetable*}

\subsection{Atomic Ions} 
We determine the total flux from [Ne II], [Ne III], and [Ar II] within the central aperture, which contains primarily emission from the disk, by fitting a Gaussian to each line. As the [Ne III] overlaps with a nearby water line, two Gaussians are fit, though only the Gaussian centered on the [Ne III] is used to determine the flux, see Section~\ref{app:atom}. 
Analysis of ions within the disk can be used to constrain the high-energy component of the stellar radiation field impinging on the disk, which in turn can drive the disk wind. In particular, the ratio of [Ne II]/[Ne III] differentiates between ionization from hard EUV photons and soft EUV and/or X-rays while [Ne II]/[Ar II] differentiates between hard X-rays and soft X-rays or EUV \citep{Hollenbach09}. However, for SY Cha the [Ne II] emission clearly includes a substantial jet component, which can be driven by locally produced UV emission from shocks \citep[e.g.,][]{Hollenbach97} and is thus not a reliable tracer of the stellar radiation field. Higher spectral resolution observations \citep[e.g.,][]{Pascucci20} would be needed to separate out the jet emission from any disk emission.

\section{Discussion}

\subsection{Extended Molecular Hydrogen Emission}

\citet{Ginski24} recently analysed scattered light observations of SY Cha at 1.6 and 2.2 \micron. They derive a disk surface height at 72 au of 7-12 au. This is significantly closer to the disk midplane than the rotational \hh\ emission, which has a height of 35 au at a radius of 72 au assuming an opening angle of 64\degree. This suggests the \hh\ does not originate from the disk surface. This can also be seen in Figure~\ref{fig:itr9}, where the scattered light surface as imaged by SPHERE is noticeably flatter than the \hh\ surface.
Additionally, our rotation diagram analysis finds an excitation temperature of $1260\pm90$~K for the lines seen in extended emission, much hotter than the expected disk warm molecular layer at large radii \citep{Law23,PanequeCarreno23}. Therefore the extended \hh\ emission is likely tracing an outflow.

Previous studies of JWST-MIRI extended \hh\ emission in similar systems report narrower semi-opening angles of 9\degree\ to 35\degree\ and similar to higher temperatures, attributed to a scattered MHD disk wind \citep{Arulanantham24,Delabrosse24}. These systems are more inclined than SY Cha and the extended \hh\ was observed in the $S$(2) and $S$(1) transitions, which do not appear extended for SY Cha. 
Interestingly, \citet{Delabrosse24} find the red-shifted wind in DG Tau B has a very narrow semi-opening angle, while the semi-opening angle of the blue-shifted emission is much wider.
\citet{Pascucci24} used JWST-NIRSpec to resolve \hh\ disk winds in the $S$(9) transition for four edge-on disks. They find semi-opening angles ranging from $13\degree$ to $50\degree$ for the blue-shifted wind. The semi-opening angle for the $S$(9) Tau 042021 is about $40\degree$, compared to the value of $40\degree$ for $S$(2) found by \citet{Arulanantham24} using MIRI observations.
Based on our derived temperature and semi-opening angle, we conclude the extended \hh\ emission in SY Cha is likely from a disk wind. This analysis of a spatially extended molecular wind toward a moderately inclined protoplantetary disk, along with several previous studies \citep[e.g.,][]{Gudel18,Booth21,Delabrosse24}, demonstrates that studies of resolved disk winds need not focus on edge-on systems.

The 600~K temperature of the warm \hh\ component is consistent with a disk origin. 
However, the derived warm \hh\ column density of $6.0 \pm 0.1 \ee{19}$~cm$^\mathrm{-2}$ is significantly smaller than the \hh\ column density of $5\ee{25}$~cm$^\mathrm{-2}$ for the very low-mass star 2MASS-J16053215-1933159 derived by \citet{Franceschi24}. This could be due to the shorter evolutionary times for disks around very low-mass stars, resulting in higher amounts of dust settling and more observable gas above the optically thick dust layer \citep{Pinilla13}. The idea that disks around low mass stars evolve more quickly has previously been used to explain the abundance of gas-phase hydrocarbons observed toward low-mass sources \citep{Tabone23,Arabhavi24}. 

\subsection{Wind Mass Loss Rate}
Assuming a typical wind velocity of 10 km\,s$\mathrm{^{-1}}$ \citep{Alexander14,Tabone20b} we can convert our derived hot \hh\ mass into a molecular wind mass loss rate, following the method of \citet{Arulanantham24}. We assume a maximum height above the midplane of 271 au based on the extent of the $S$(5) emission in Figure~\ref{fig:itr9} and assume an inclination angle of 51.1\degree. This results in a wind mass loss rate of $3\pm2\ee{-9}$ \msun\ yr$^\mathrm{-1}$, an order of magnitude greater than the mass accretion rate onto the central star reported in Table~\ref{sourcetab} and in line with most reported molecular wind mass loss rates \citep{Louvet18,deValon20,Tabone20b,Arulanantham24}.
We note that SY Cha is a transition disk with a low mass accretion rate. The disk is presumably more evolved than previously characterized systems, which could explain the lower wind mass loss rate. Additionally, the molecular wind component is likely only a fraction of the total mass in the wind, with a significant fraction of the mass in neutral and ionized atomic gas \citep{Nakatani18,Wang19,Komaki21,Sellek24b}. 
However, our derived wind mass loss rate far exceeds the typical $\dot{M}_{wind}/\dot{M}_{acc}$ of 0.1-1 \citep{Pascucci23}.
As MHD winds are driven by accretion energy while photoevaporative winds are thermally driven, and thus do not transport significant angular momentum, \citep{Lesur24}, it is possible that the extended \hh\ emission is from a photoevaporative wind, which does not relate to stellar accretion, though this would not explain the [Ne II] emission. Another possibility is that SY Cha's stellar accretion rate is highly variable and was measured at a particularly low point. This would also explain the mid-IR variability between the three epochs of \spitzer\ observations and the \jwst\ observations reported by \citet{Schwarz24}.


\subsection{Extended [Ne II] Emission}
[Ne II] emission is spatially resolved, with both a blue shifted and red shifted component. The [Ne II] emission is both more extended and more collimated than the extended \hh\ emission. Similar results from JWST-MIRI were recently reported for the edge-on source Tau 042021, where the jet was also seen in [Ne III], [Ni II], [Fe II], [Ar II], and [S III] \citep{Arulanantham24}. These lines do not appear to be spatially extended in our observations of SY Cha, though this may be due in part to the weaker line emission. The undetected lines such as [Ni II], [Fe II], and [S III] could also be masked by stronger water lines in the central region.  

The [Ne II] emission with a small velocity offset appears largely unresolved and co-spatial with the central source, while there are also clearly two lobes of emission extending away from the disk corresponding to velocity offsets of greater than 100 km\,s$^{-1}$. High spectral resolution observations of [Ne II] toward other sources often show either a high-velocity component tracing a jet for sources with high accretion rates, or a low-velocity component tracing a wind for low accretors  \citep[e.g.,][]{Pascucci20}. Recent \emph{JWST} MIRI MRS observations of [Ne II] in edge-on disks are likewise either attributed to a disk wind \citep[e.g.,][]{Bajaj24} or a jet \citep{Arulanantham24}.
For SY Cha, only the [Ne II] emission with a large velocity offset appears spatially extended.
As the [NeII] emission is extended and collimated, the wind origin seems to be excluded.
This nested morphology, with \hh\ tracing a wide angle wind and atomic forbidden line emission tracing a more collimated jet has been seen in several sources \citep{Pascucci24}.
We conclude that while there may be a spatially unresolved wind component to the [Ne II] emission close to the central star, the dominant extended component comes from a jet. SY Cha is unusual in that it appears to be is a low accretor with a [Ne II] jet.

\section{Conclusions}
We report spatially extended emission surrounding SY Cha observed with MIRI-MRS in the pure rotational $S$(3)-$S$(7) \hh\ lines and [Ne II], as well as unresolved detections of [Ne III] and [Ar II]. Our conclusions are as follows:
\begin{itemize}[noitemsep]
\item The [Ne II] emission is highly collimated, with both blue and red-shifted lobes. This, as well as a narrow semi-opening angle of $6.3\pm1.8\degree$ is indicative of a jet. 
\item The \hh\ emission is seen primarily on the near side of the source. 
\item The semi-opening angle of the \hh\ is much wider than the [Ne II], ranging from $48\pm1$ to $64\pm3\degree$ depending on the line. 
\item The \hh\ rotation diagram is best described by a two component fit. The lines for which we see extended emission are all fit by a temperature of $1260\pm 90$~K and a column density of $1.1  \pm  0.3 \ee{18}$~cm$^{-2}$, while the lower excitation temperature lines are fit by a temperature of $600  \pm  100$ and a column density of $2  \pm  1 \ee{19}$~~cm$^{-2}$. 
\item Based on the wide semi-opening angle, high temperature, and low column density for the extended \hh, we conclude it originates from a disk wind. 
\item We derive a molecular wind mass loss rate of $3\pm2\ee{-9}$ \msun\ yr$^\mathrm{-1}$, in line with most literature value but unexpectedly high compared to the previously derived stellar accretion rate of $6.6\ee{-10}$ \msun\ yr$^\mathrm{-1}$. This points to either a photoevaporative origin for the wind or variable accretion onto the central star.
\end{itemize}

\section{Software \& Facilities}

\facility{JWST}
\software{\textsc{emcee} \citep{emcee}, \textsc{matplotlib} \citep{matplotlib}, \textsc{MINDS pipeline} \citep{mindspipeline},\textsc{numpy} \citep{numpy}, \textsc{spectres} \citep{Carnall17}, \textsc{VIP} \citep{VIP1,VIP2}}

\section{Acknowledgements}

This work is based on observations made with the NASA/ESA/CSA James Webb Space Telescope. The data were obtained from the Mikulski Archive for Space Telescopes at the Space Telescope Science Institute, which is operated by the Association of Universities for Research in Astronomy, Inc., under NASA contract NAS 5-03127 for JWST. 
These observations are associated with program 1282. The following National and International Funding Agencies funded and supported the MIRI development: 
NASA; ESA; Belgian Science Policy Office (BELSPO); Centre Nationale d’Etudes Spatiales (CNES); Danish National Space Centre; Deutsches Zentrum fur Luft und Raumfahrt (DLR); Enterprise Ireland; 
Ministerio De Econom\'ia y Competividad; 
Netherlands Research School for Astronomy (NOVA); 
Netherlands Organisation for Scientific Research (NWO); 
Science and Technology Facilities Council; Swiss Space Office; 
Swedish National Space Agency; and UK Space Agency.

The data presented in this article were obtained from the Mikulski Archive for Space Telescopes (MAST) at the Space Telescope Science Institute. The specific observations analyzed can be accessed via \dataset[doi: 10.17909/61c6-ah42]{https://doi.org/10.17909/61c6-ah42}.

K.S. and T.H. acknowledge support from the European Research Council under the Horizon 2020 Framework Program via the ERC Advanced Grant Origins 83 24 28. 
B.T. is a Laureate of the Paris Region fellowship program, which is supported by the Ile-de-France Region and has received funding under the Horizon 2020 innovation framework program and Marie Sklodowska-Curie grant agreement No. 945298.
M.T., M.V. and A.D.S acknowledge support from the ERC grant 101019751 MOLDISK.
I.K., A.M.A., and E.v.D. acknowledge support from grant TOP-1 614.001.751 from the Dutch Research Council (NWO). 

I.K., J.K., and T.K. acknowledge funding from H2020-MSCA-ITN-2019, grant no. 860470 (CHAMELEON).
E.v.D. acknowledges support from the ERC grant 101019751 MOLDISK and the Danish National Research Foundation through the Center of Excellence ``InterCat'' (DNRF150). 
A.C.G. acknowledges support from PRIN-MUR 2022 20228JPA3A “The path to star and planet formation in the JWST era (PATH)” funded by NextGeneration EU and by INAF-GoG 2022 “NIR-dark Accretion Outbursts in Massive Young stellar objects (NAOMY)” and Large Grant INAF 2022 “YSOs Outflows, Disks and Accretion: towards a global framework for the evolution of planet forming systems (YODA)”.
T.P.R acknowledges support from ERC grant 743029 EASY.
V.C. acknowledgew funding from the Belgian F.R.S.-FNRS.
D.G. thanks the Belgian Federal Science Policy Office (BELSPO) for the provision of financial support in the framework of the PRODEX Programme of the European Space Agency (ESA).
G.P. gratefully acknowledges support from the Max Planck Society.

\begin{appendix}
\section{Image De-convolution}\label{app:deconvolution}
\begin{figure*}
    \centering
    \includegraphics[width=0.9\textwidth]{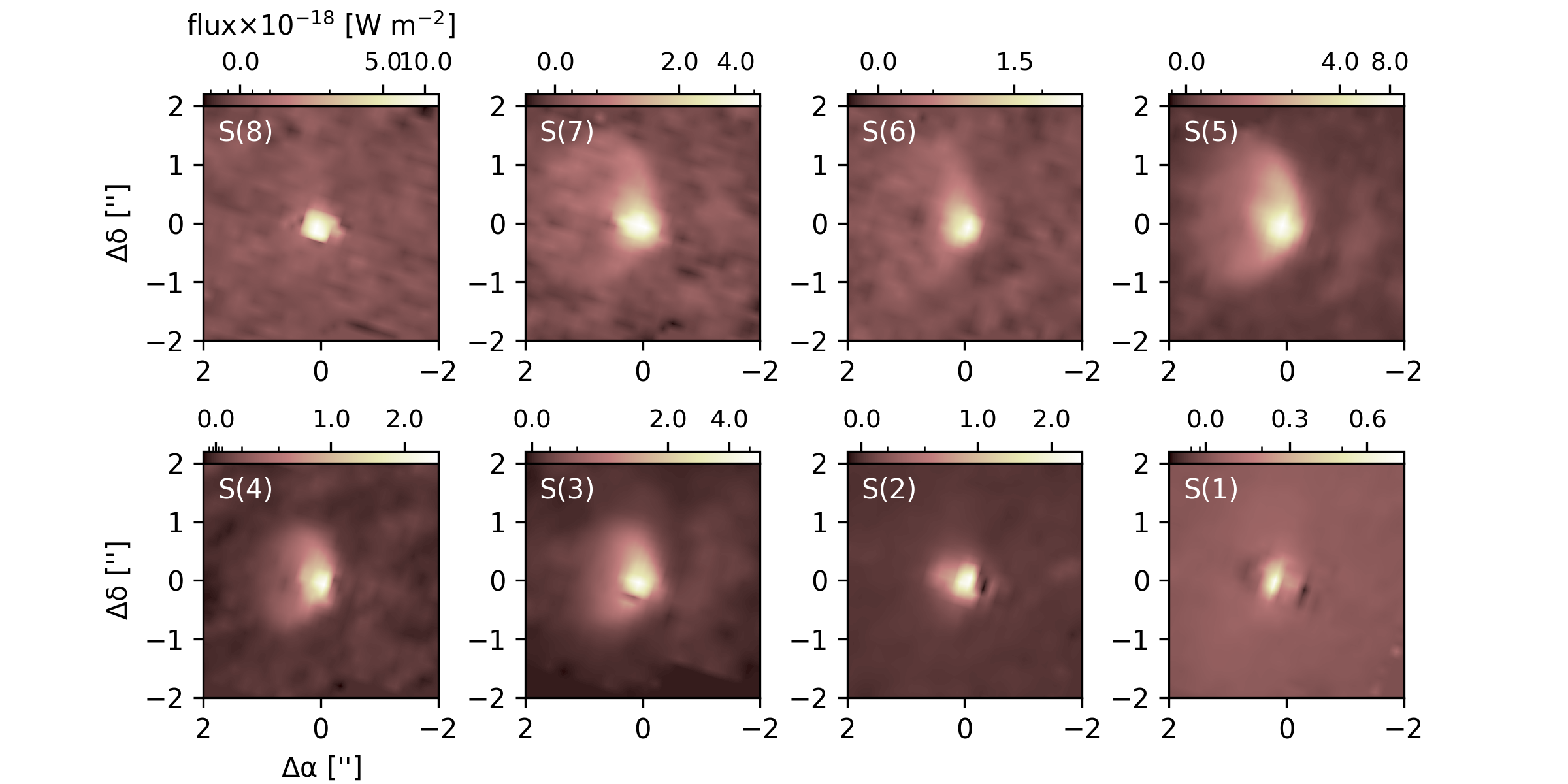}
    \caption{\hh\ lines after PSF subtraction but before de-convolution.}
    \label{fig:itr0}
\end{figure*}

\begin{figure*}
    \centering
    \includegraphics[width=0.9\textwidth]{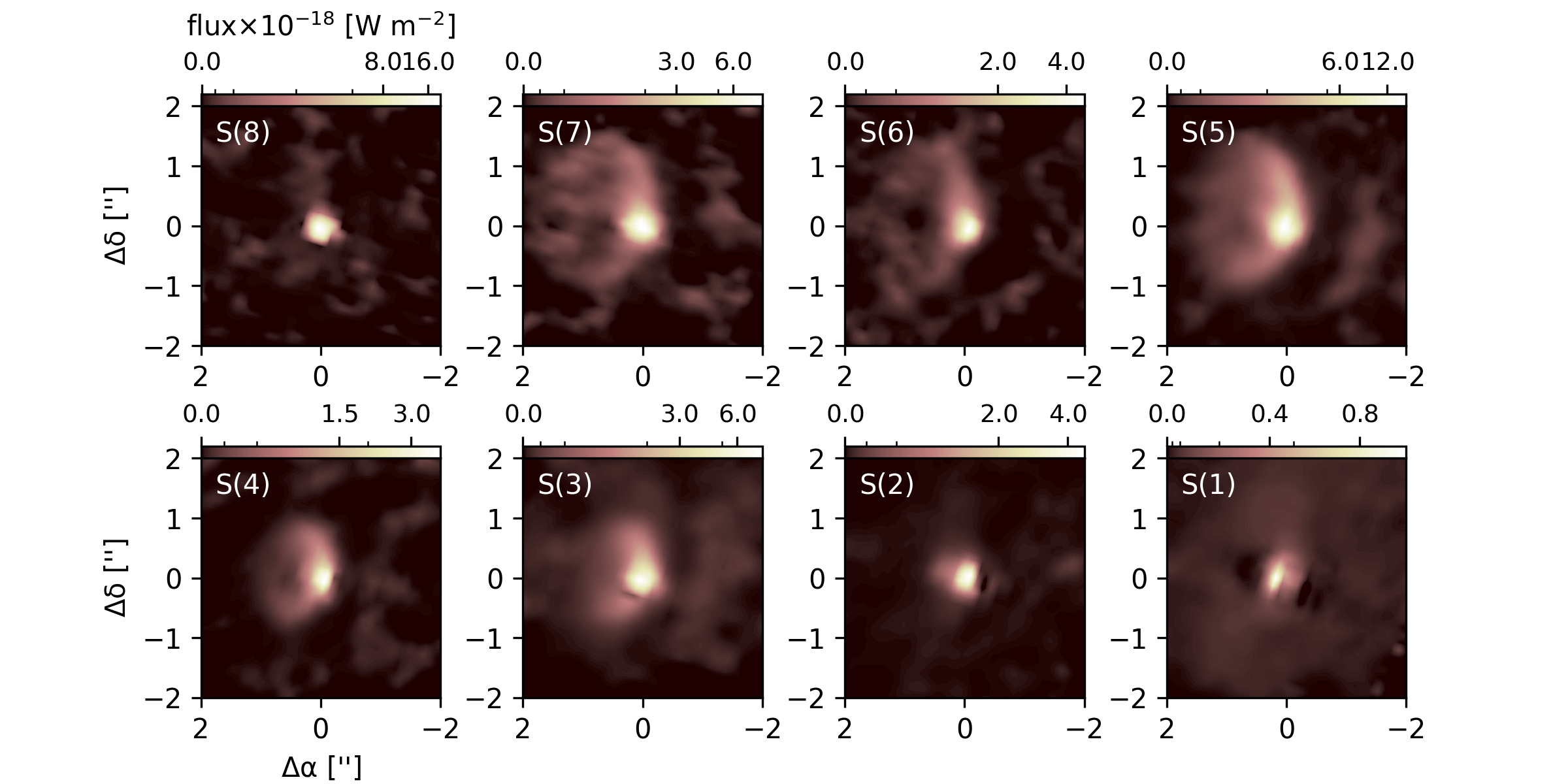}
    \caption{\hh\ lines after three rounds of de-convolution.}
    \label{fig:itr3}
\end{figure*}

\begin{figure*}
    \centering
    \includegraphics[width=0.9\textwidth]{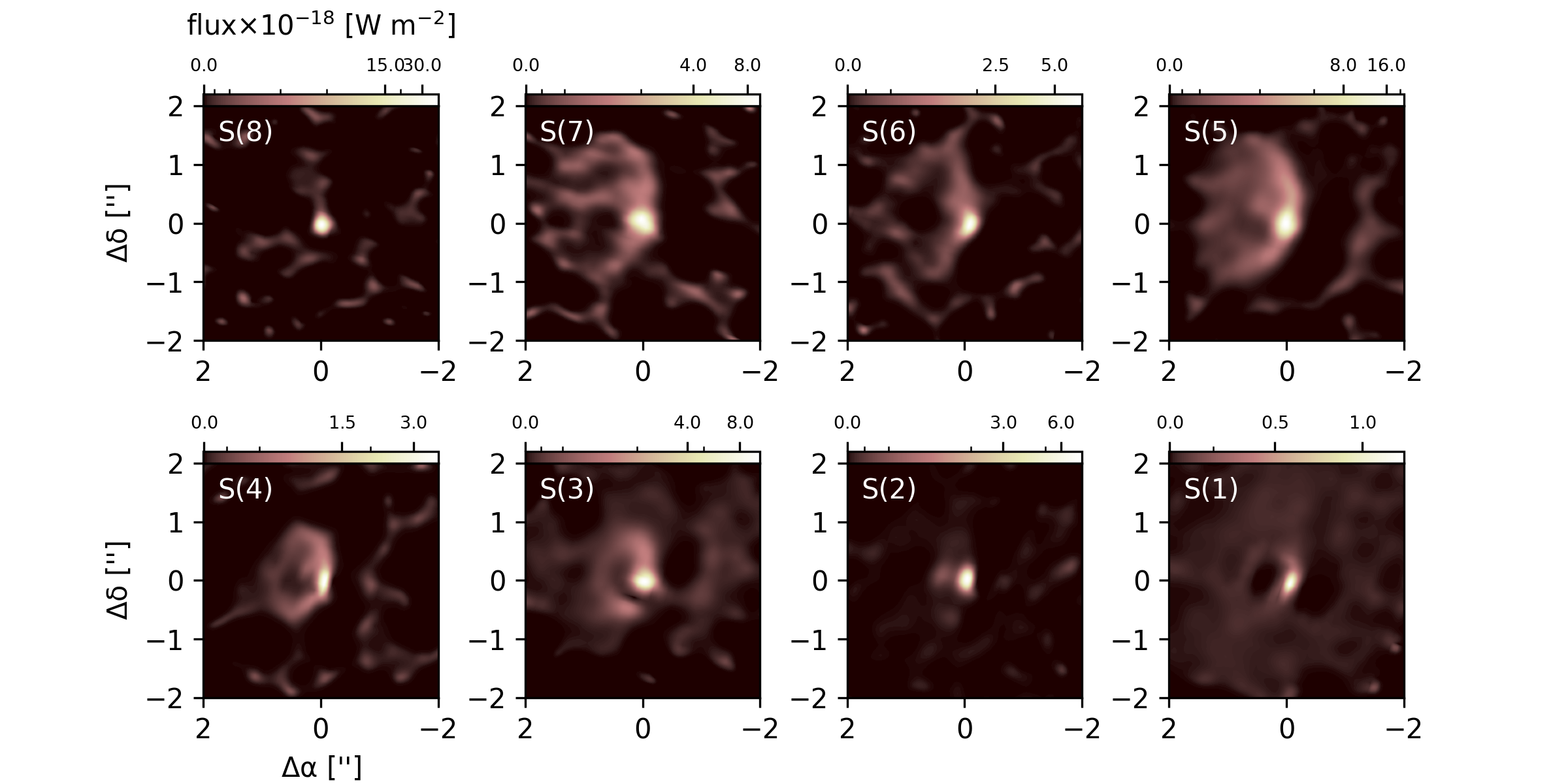}
    \caption{\hh\ lines after 27 rounds of de-convolution.}
    \label{fig:itr27}
\end{figure*}

Here we present additional \hh\ images: after PSF subtraction but before de-convolution (Figure~\ref{fig:itr0}), after three rounds of de-convolution (Figure~\ref{fig:itr3}), and after 27 rounds (Figure~\ref{fig:itr27}).

\section{Channel Maps}\label{app:channel}
Here we present the channel maps for lines with extended emission Figures~\ref{fig:NeIIchan}. These maps have been continuum subtracted using the flux in adjacent channels. No de-convolution has been applied. 

\begin{figure}
    \centering
    \includegraphics{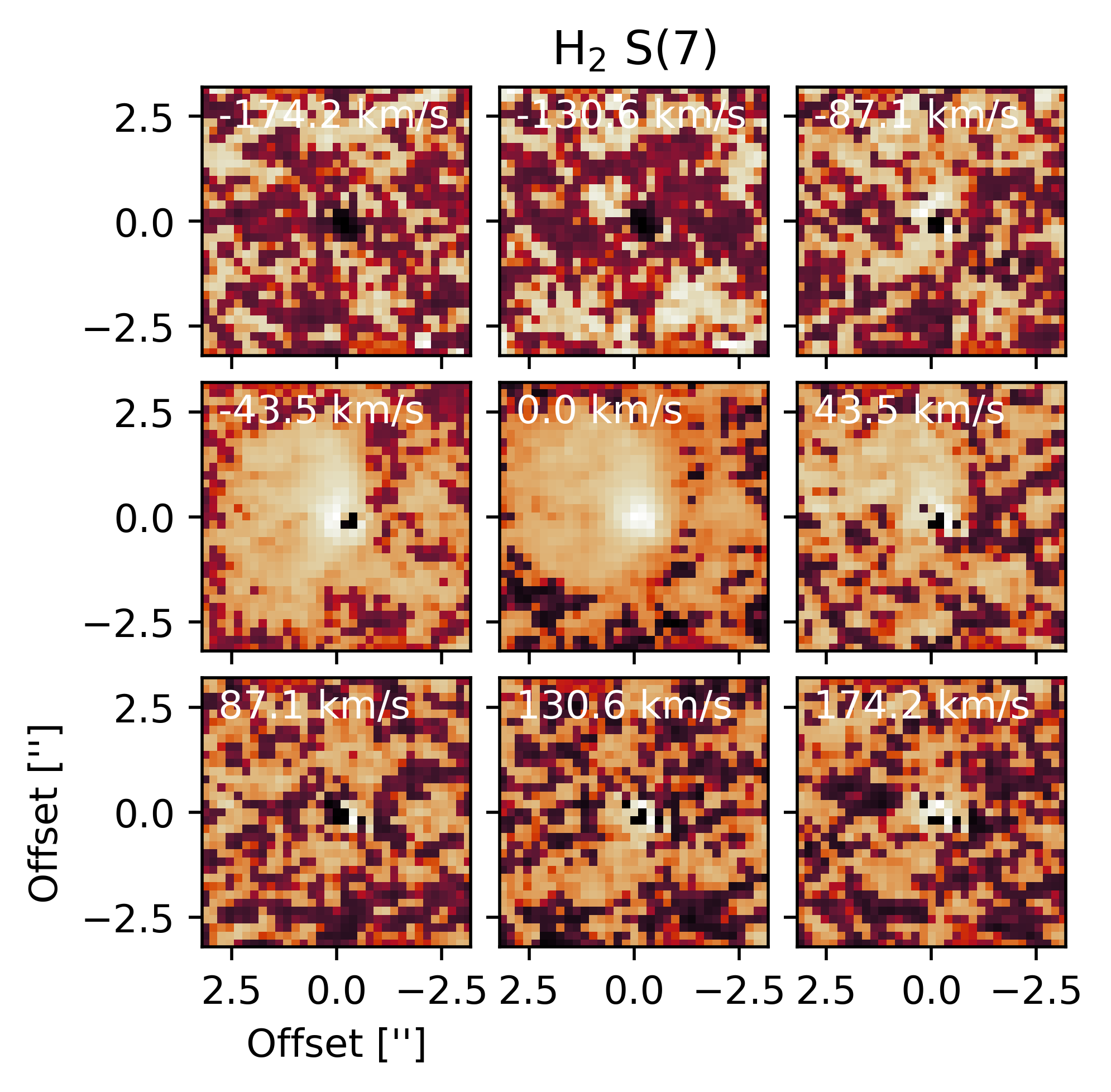}
    \caption{Channel map of the continuum subtracted data cube for channels surround the \hh\ S(7) line. Velocity offset from the line center are shown in the top left of each pannel.}
    \label{fig:H27chan}
\end{figure}

\begin{figure}
    \centering
    \includegraphics{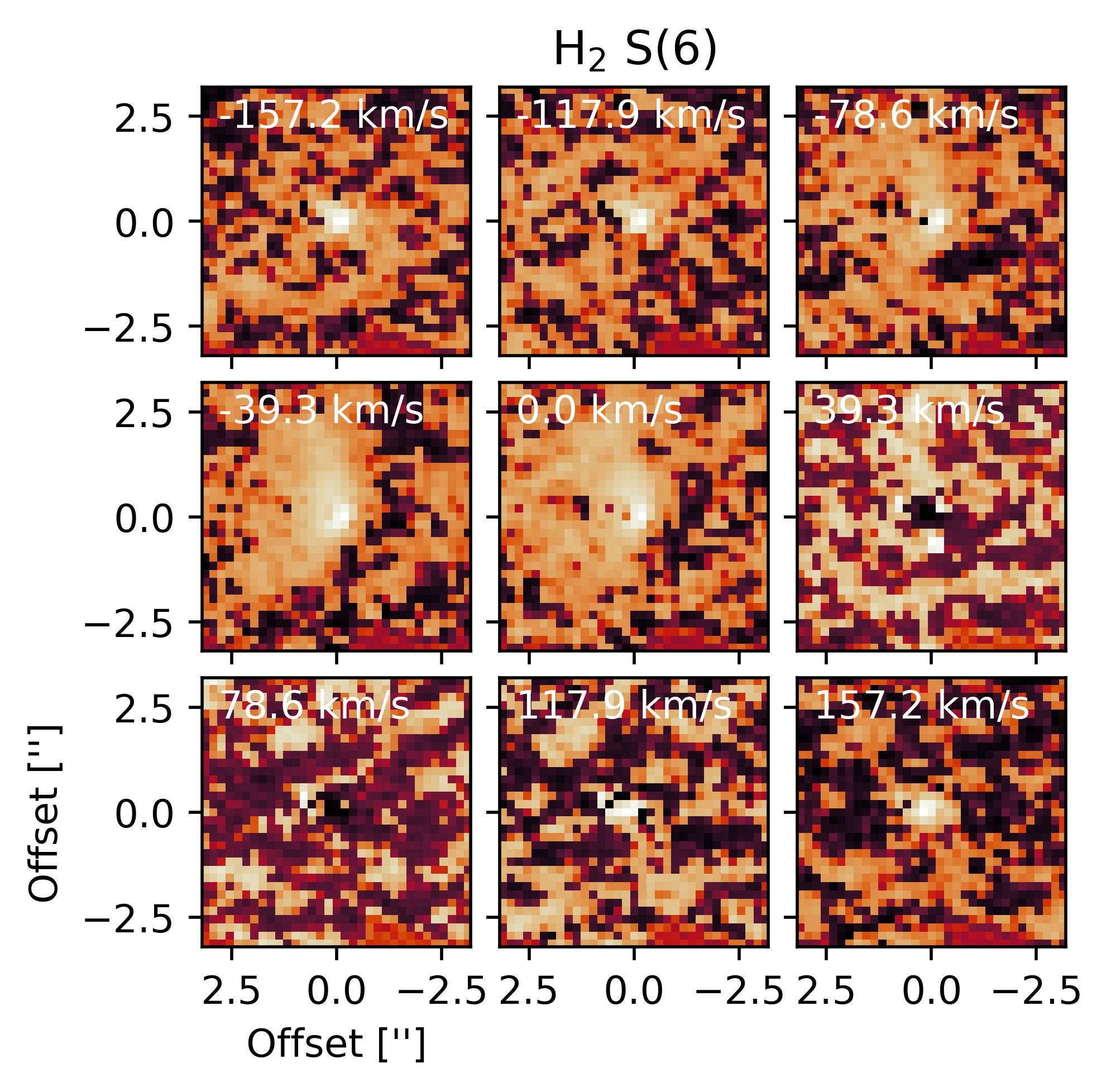}
    \caption{Same as Figure \ref{fig:H27chan} but for \hh\ S(6).}
    \label{fig:H26chan}
\end{figure}

\begin{figure}
    \centering
    \includegraphics{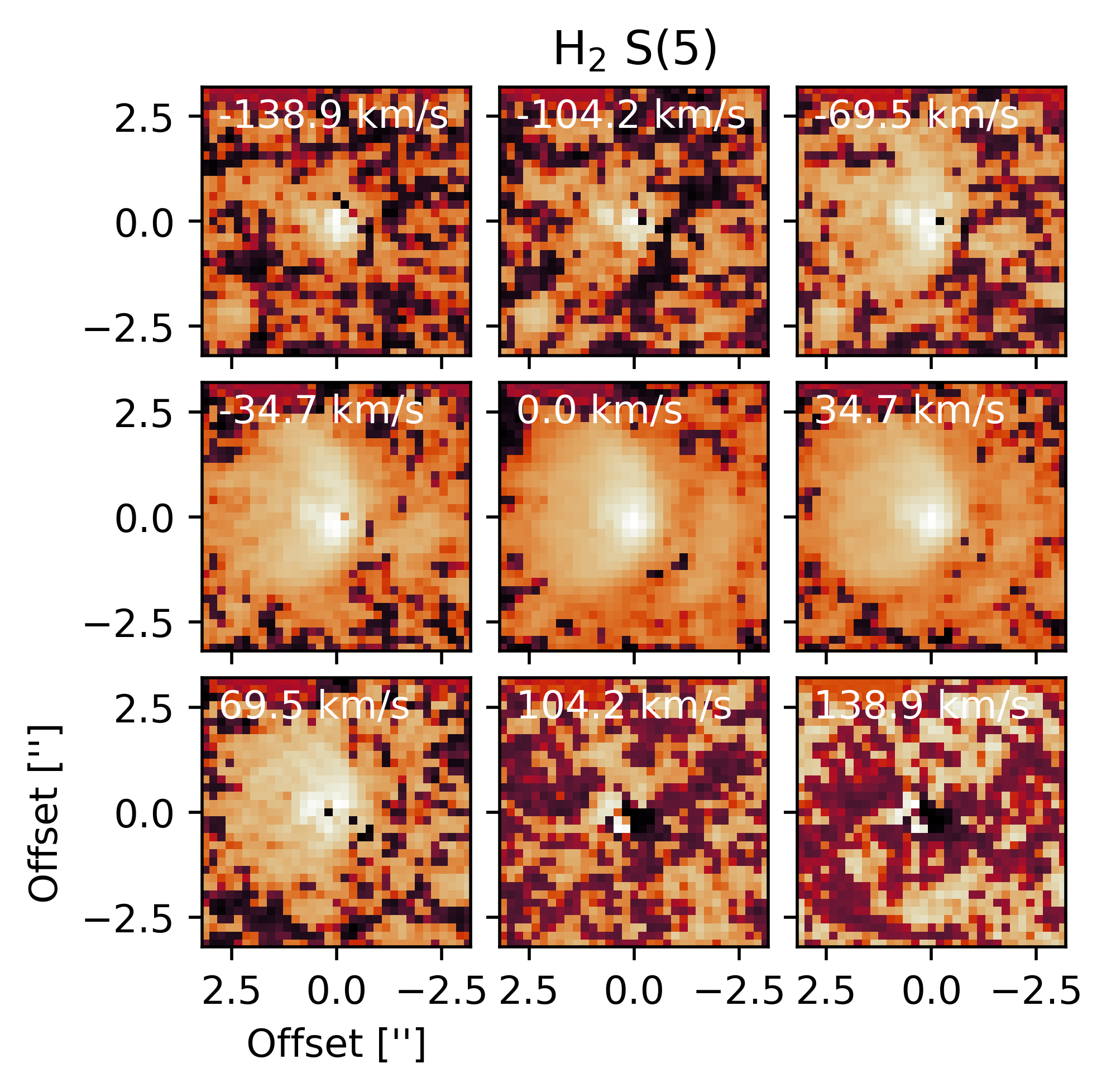}
    \caption{Same as Figure \ref{fig:H27chan} but for \hh\ S(5).}
    \label{fig:H25chan}
\end{figure}

\begin{figure}
    \centering
    \includegraphics{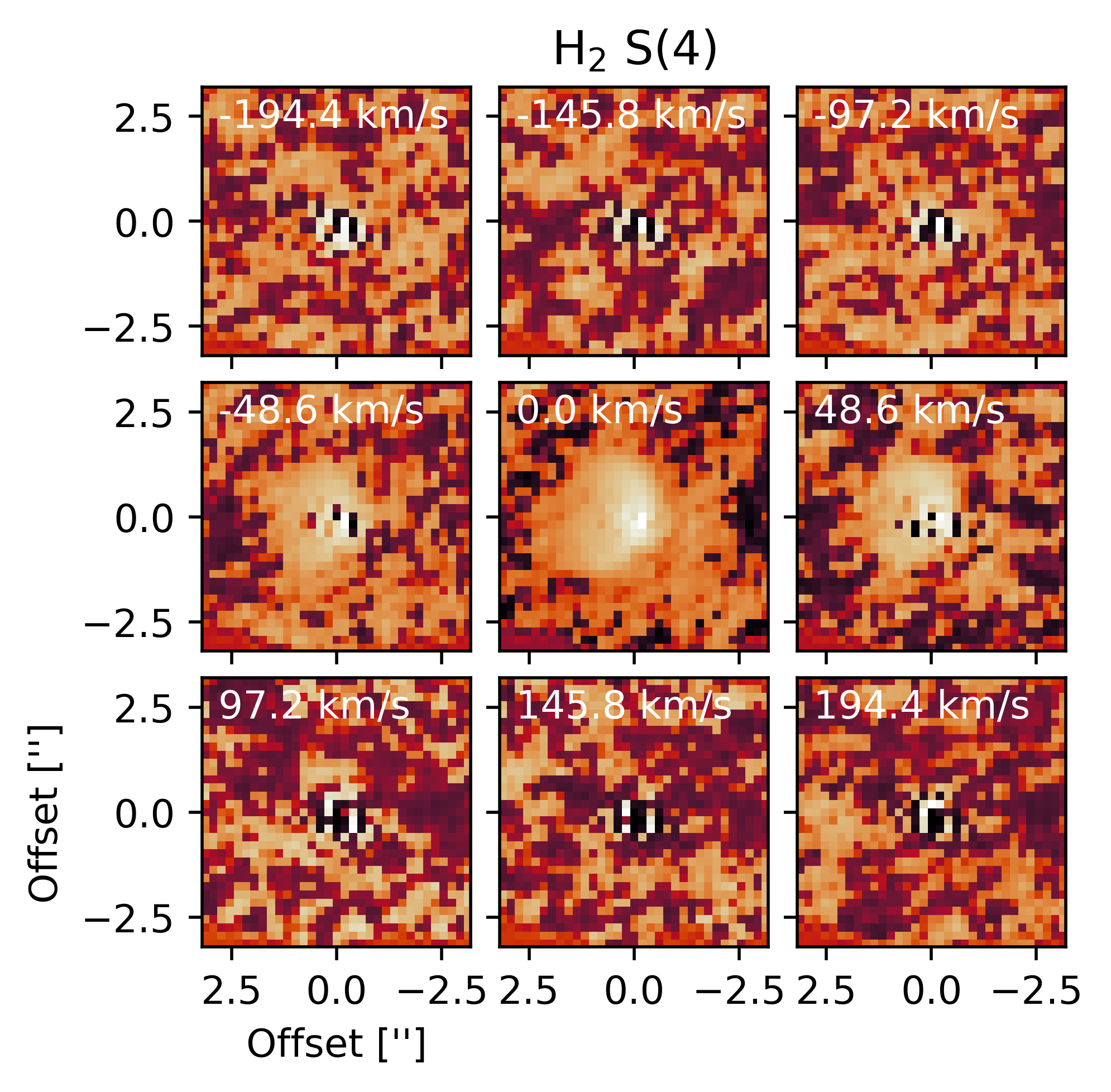}
    \caption{Same as Figure \ref{fig:H27chan} but for \hh\ S(4).}
    \label{fig:H24chan}
\end{figure}

\begin{figure}
    \centering
    \includegraphics{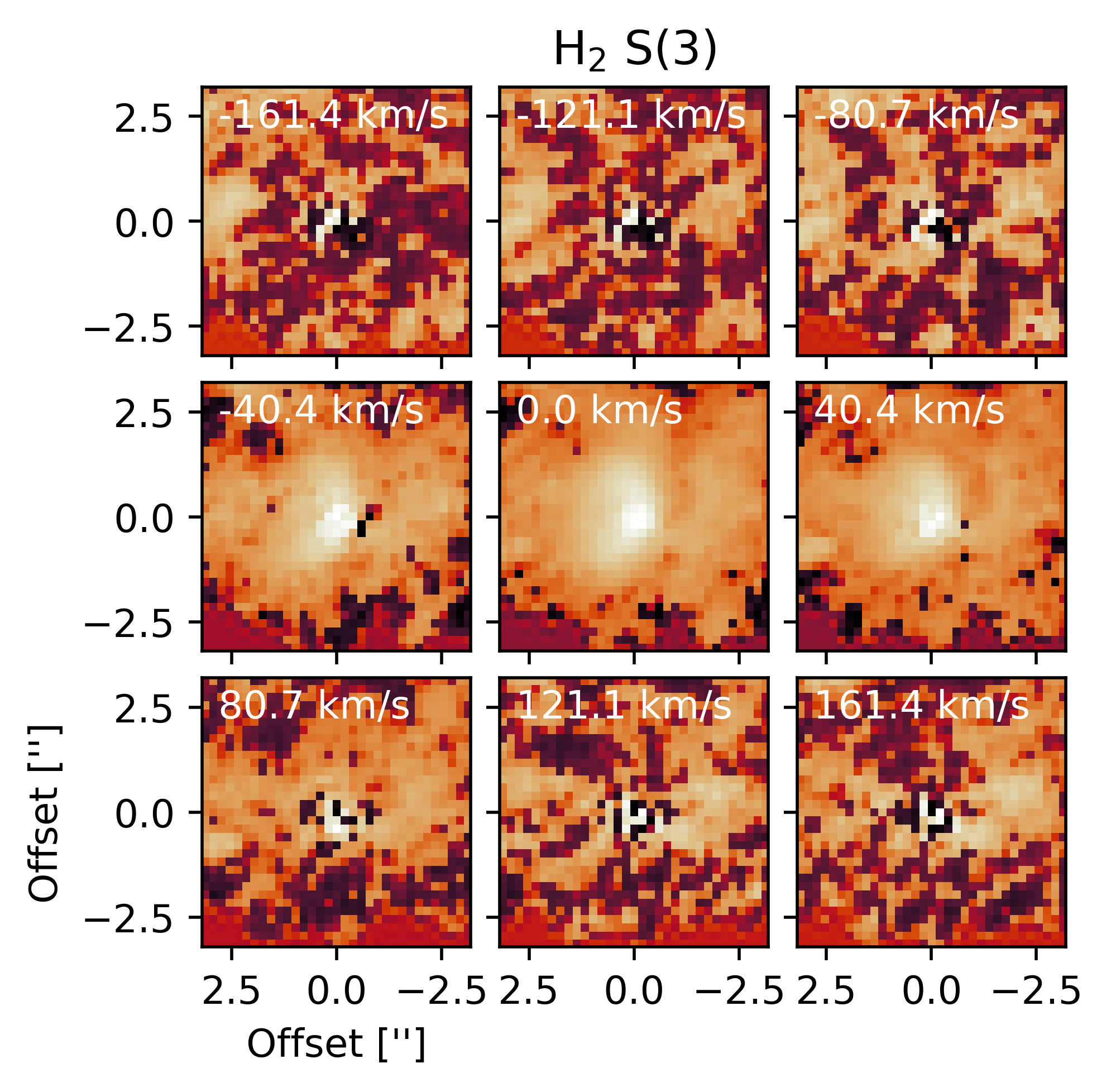}
    \caption{Same as Figure \ref{fig:H27chan} but for \hh\ S(3).}
    \label{fig:H23chan}
\end{figure}

\begin{figure}
    \centering
    \includegraphics{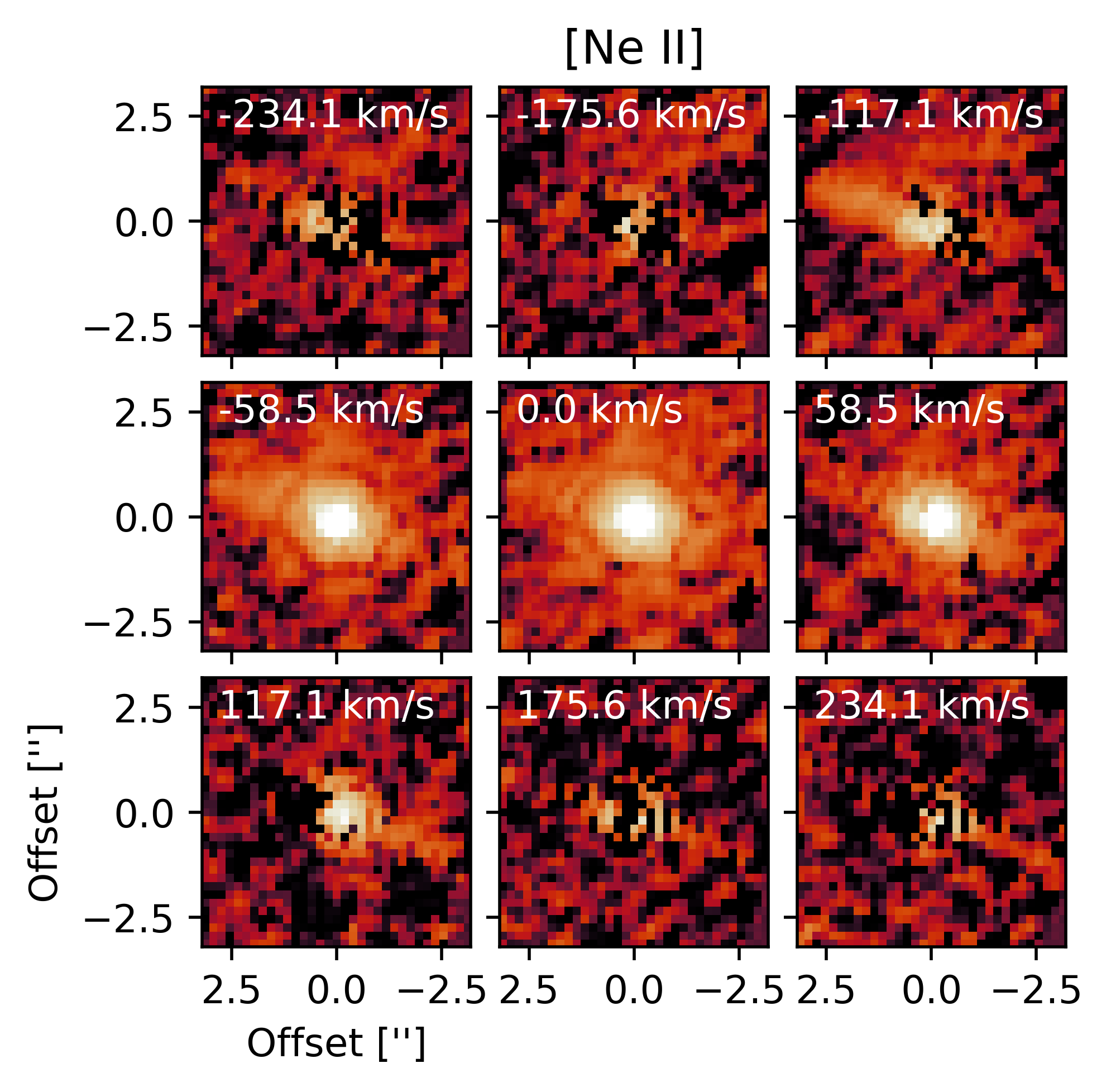}
    \caption{Same as Figure \ref{fig:H27chan} but for [Ne II]. The diffraction pattern seen near 0 km/s is due to unresolved [Ne II] emission which does not follow the subtracted continuum.}
    \label{fig:NeIIchan}
\end{figure}

\section{Fits to Atomic Lines}\label{app:atom}
\begin{figure}
    \centering
    \includegraphics{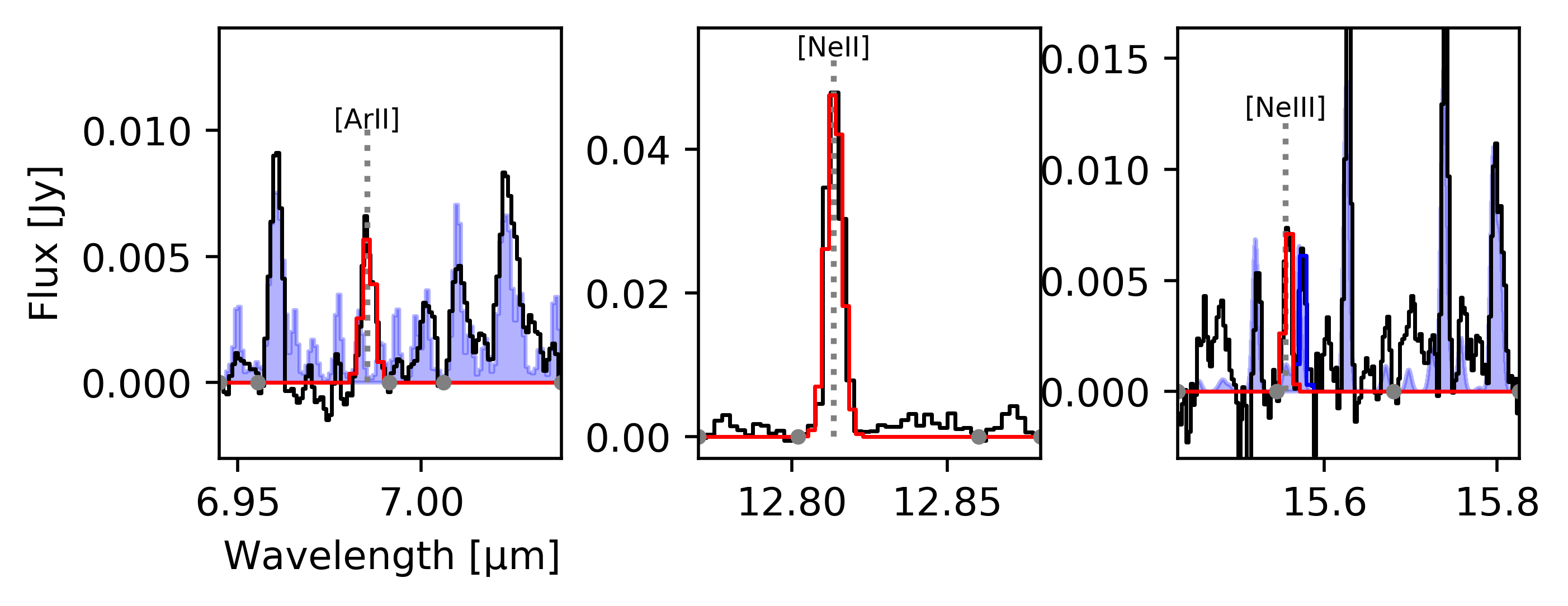}
    \caption{Best-fit solution for the Gaussians fit to the atomic lines. Black is the MRS spectrum after continuum subtraction. Grey dots show the points used for continuum subtraction. Red line is the best fit. In the right panel, the dark blue line is the best fit to the nearby water line, while the light blue is the best fit water slab model of \citet{Schwarz24}.}
    \label{fig:atoms}
\end{figure}

Here we present the results of our atomic line fits. The local continuum is determined by selecting line-free pixels and interpolating between them using a spline fit. We then fit a Gaussian line profile to the atomic line using the Markov Chain Monte Carlo package \texttt{emcee} \citep{emcee}. The best-fit models are shown in Figure~\ref{fig:atoms} and the values are given in Table~\ref{tab:atoms}.

\begin{deluxetable*}{ccccccc}
\tablewidth{0pt}
\tablecolumns{2}
\tablecaption{Integrated flux for Gaussian fits to atomic lines}
\label{tab:atoms}
\tablehead{
\colhead{Transition} & \colhead{Line Center} & \colhead{$\mathrm{\sigma}$} & \colhead{$E_u$} & \colhead{$n_{ecrit}$}   & \colhead{Integrated Flux} & \colhead{rms}  \\  
 & \colhead{[\um]} & \colhead{10$^{-3}$[\um]} & \colhead{[K]} & \colhead{[cm$^{-2}$]} & \colhead{10$^{-15}$[erg s$^{-1}$ cm$^{-2}$]} & \colhead{$10^{-28}$[erg s$^{-1}$ cm$^{-2}$ Hz$^{-1}$]} 
}
\startdata 
Ar II & 6.985  & 1.7 & 2059.8 & & 4.2e5   & 6.7  \\
Ne II & 12.814 & 2.6 & 1122.0 & 6.3e5 & 5.87  & 8.5   \\
Ne III & 15.555 & 4.0 & 924.9 & 2.7e5 & 0.94 & 9.0   \\
\enddata 

\end{deluxetable*}

\end{appendix}

\bibliography{manuscript}{}
\bibliographystyle{aasjournal}

\end{document}